\documentclass[12pt,a4paper]{article}
\makeatletter
\def\eqnarray{%
\stepcounter{equation}%
\let\@currentlabel=\theequation
\global\@eqnswtrue
\global\@eqcnt\z@
\tabskip\@centering
\let\\=\@eqncr
$$\halign to \displaywidth\bgroup\@eqnsel\hskip\@centering
$\displaystyle\tabskip\z@{##}$&\global\@eqcnt\@ne
\hfil$\displaystyle{{}##{}}$\hfil
&\global\@eqcnt\tw@$\displaystyle\tabskip\z@{##}$\hfil
\tabskip\@centering&\llap{##}\tabskip\z@\cr}
\makeatother
\newcommand{\ket}[1]{{\vert{#1}\rangle}}

\newcommand{\calh}{{\cal H}}

\newcommand{\fukuso}{{\mathbf C}}

\begin{document}

\title{\sl Quantum Diagonalization Method in \\
the Tavis--Cummings Model}
\author{
  Kazuyuki FUJII
  \thanks{E-mail address : fujii@yokohama-cu.ac.jp },\quad 
  Kyoko HIGASHIDA 
  \thanks{E-mail address : s035577d@yokohama-cu.ac.jp },\quad 
  Ryosuke KATO 
  \thanks{E-mail address : s035559g@yokohama-cu.ac.jp }\\
  Tatsuo SUZUKI
  \thanks{E-mail address : suzukita@gm.math.waseda.ac.jp },\quad 
  Yukako WADA 
  \thanks{E-mail address : s035588a@yokohama-cu.ac.jp }\\
  ${}^{*,\dagger,\ddagger,\P}$Department of Mathematical Sciences\\
  Yokohama City University, 
  Yokohama, 236--0027, 
  Japan\\
  ${}^\S$Department of Mathematical Sciences\\
  Waseda University, 
  Tokyo, 169--8555, 
  Japan\\
  }
\date{}
\maketitle
%
%
%
%
\begin{abstract}
  To obtain the explicit form of evolution operator in the Tavis--Cummings 
  model we must calculate the term 
  $\mbox{e}^{-itg(S_{+}\otimes a+S_{-}\otimes a^{\dagger})}$ explicitly 
  which is very hard. In this paper we try to make the quantum matrix 
  $A\equiv S_{+}\otimes a+S_{-}\otimes a^{\dagger}$ diagonal to calculate 
  $\mbox{e}^{-itgA}$ and, moreover, to know a deep structure of the model.
  
  For the case of one, two and three atoms we give such a diagonalization 
  which is first nontrivial examples as far as we know, and reproduce 
  the calculations of $\mbox{e}^{-itgA}$ given in quant-ph/0404034. We 
  also give a hint to an application to a noncommutative differential 
  geometry. 
  
  However, a quantum diagonalization is not unique and is affected by some 
  ambiguity arising from the noncommutativity of operators in quantum physics. 
  
  Our method may open a new point of view in Mathematical Physics or 
  Quantum Physics. 
\end{abstract}
%


%
%
%
%

\section{Introduction}
The purpose of this paper is to give a new insight to the Tavis--Cummings 
model (\cite{TC}) and to obtain the explicit form of evolution operator 
by the new method in the case of some atoms. 

This model is a very important one in Quantum Optics and (maybe even) in 
Mathematical Physics, and has been studied widely, see \cite{books} as 
general textbooks in quantum optics. 

We are studying a quantum computation and therefore want to study the model 
from this point of view, namely the quantum computation based on atoms of 
laser--cooled and trapped linearly in a cavity. We must in this model 
construct the controlled NOT gate or other controlled unitary gates to perform 
the quantum computation, see \cite{elementary-gate} as a general introduction 
to this subject. 

For that aim we need the explicit form of evolution operator of the model 
in the case of (at least) one, two and three atoms. 
As to the model of one atom or two atoms it is more or less known (see 
\cite{papers-1}), while as to the case of three atoms it was given by 
\cite{papers-2}. However, the method is not clear enough in a mathematical 
sense \footnote{We used Mathematica in the process of calculation}. 

In this paper we present a quantum diagonalization method 
which is a quantum version of classical diagonalization and obtain 
the explicit form of evolution operator obtained in \cite{papers-2}. 

However, the quantum diagonalization is not unique and is affected by some 
ambiguity due to the noncommutativity of operators in Quantum Physics. 
This may be related to the so--called operator ordering problem, see for 
example \cite{FS} on this topics.

The Quantum Diagonalization Method is completely new and may be applied to 
a noncommutative differential geometry (for example, the noncommutative 
chiral models) because we can construct (quantum) unitary matrices explicitly. 
However, this is beyond our scope of the paper.

\section{Tavis--Cummings Model and Evolution Operator}

We make a review of \cite{papers-2} within our necessity. 
The Tavis--Cummings model (with $n$--atoms) that we will treat in this paper 
can be written as follows (we set $\hbar=1$ for simplicity). 
\begin{equation}
\label{eq:hamiltonian}
H=
\omega {1}_{L}\otimes a^{\dagger}a + 
\frac{\Delta}{2} \sum_{i=1}^{n}\sigma^{(3)}_{i}\otimes {\bf 1} +
g\sum_{i=1}^{n}\left(
\sigma^{(+)}_{i}\otimes a+\sigma^{(-)}_{i}\otimes a^{\dagger} \right),
\end{equation}
where $\omega$ is the frequency of radiation field, $\Delta$ the energy 
difference of two level atoms, $a$ and $a^{\dagger}$ are 
annihilation and creation operators of the field, and $g$ a coupling constant, 
and $L=2^{n}$. Here $\sigma^{(+)}_{i}$, $\sigma^{(-)}_{i}$ and 
$\sigma^{(3)}_{i}$ are given as 
\begin{equation}
\sigma^{(s)}_{i}=
1_{2}\otimes \cdots \otimes 1_{2}\otimes \sigma_{s}\otimes 1_{2}\otimes \cdots 
\otimes 1_{2}\ (i-\mbox{position})\ \in \ M(L,\fukuso)
\end{equation}
where $s$ is $+$, $-$ and $3$ respectively and 
\begin{equation}
\label{eq:sigmas}
\sigma_{+}=
\left(
  \begin{array}{cc}
    0& 1 \\
    0& 0
  \end{array}
\right), \quad 
\sigma_{-}=
\left(
  \begin{array}{cc}
    0& 0 \\
    1& 0
  \end{array}
\right), \quad 
\sigma_{3}=
\left(
  \begin{array}{cc}
    1& 0  \\
    0& -1
  \end{array}
\right), \quad 
1_{2}=
\left(
  \begin{array}{cc}
    1& 0  \\
    0& 1
  \end{array}
\right).
\end{equation}

Here let us rewrite the hamiltonian (\ref{eq:hamiltonian}). If we set 
\begin{equation}
\label{eq:large-s}
S_{+}=\sum_{i=1}^{n}\sigma^{(+)}_{i},\quad 
S_{-}=\sum_{i=1}^{n}\sigma^{(-)}_{i},\quad 
S_{3}=\frac{1}{2}\sum_{i=1}^{n}\sigma^{(3)}_{i},
\end{equation}
then (\ref{eq:hamiltonian}) can be written as 
\begin{equation}
\label{eq:hamiltonian-2}
H=
\omega {1}_{L}\otimes a^{\dagger}a + \Delta S_{3}\otimes {\bf 1} + 
g\left(S_{+}\otimes a + S_{-}\otimes a^{\dagger} \right)
\equiv H_{0}+V,
\end{equation}
which is very clear. We note that $\{S_{+},S_{-},S_{3}\}$ satisfy the 
$su(2)$--relation 
\begin{equation}
[S_{3},S_{+}]=S_{+},\quad [S_{3},S_{-}]=-S_{-},\quad [S_{+},S_{-}]=2S_{3}.
\end{equation}
However, the representation $\rho$ defined by 
$
\rho(\sigma_{+})=S_{+},\ \rho(\sigma_{-})=S_{-},\ 
\rho(\sigma_{3}/2)=S_{3}
$
is a reducible representation of $su(2)$. 

We would like to solve the Schr{\" o}dinger equation 
\begin{equation}
\label{eq:schrodinger}
i\frac{d}{dt}U=HU=\left(H_{0}+V\right)U, 
\end{equation}
where $U$ is a unitary operator (called the evolution operator). 
We can solve this equation by using the {\bf method of constant variation}. 
The result is well--known to be 
\begin{equation}
\label{eq:full-solution}
U(t)=\left(\mbox{e}^{-it\omega S_{3}}\otimes 
\mbox{e}^{-it\omega a^{\dagger}a}\right)
\mbox{e}^{-itg\left(S_{+}\otimes a + S_{-}\otimes a^{\dagger}\right)}
\end{equation}
under the resonance condition $\Delta=\omega$, 
where we have dropped the constant unitary operator for simplicity. 
Therefore 
we have only to calculate the term (\ref{eq:full-solution}) explicitly, 
which is however a very hard task \footnote{The situation is very similar to 
that of the paper quant-ph/0312060 in \cite{qudit-papers}}. 
In the following we set 
\begin{equation}
\label{eq:A}
A=S_{+}\otimes a + S_{-}\otimes a^{\dagger}
\end{equation}
for simplicity. 
We can determine\ $\mbox{e}^{-itgA}$\ for $n=1$ (one atom case), 
$n=2$ (two atoms case) and $n=3$ (three atoms case) completely. 

\vspace{3mm}
\par \noindent 
{\bf One Atom Case}\quad In this case $A$ in (\ref{eq:A}) is written as 
\begin{equation}
\label{eq:A-one}
A_{1}=
\left(
  \begin{array}{cc}
    0&           a \\
    a^{\dagger}& 0
  \end{array}
\right)\equiv B_{1/2}.
\end{equation}
By making use of the relation 
\begin{equation}
\label{eq:relation-one}
{A_{1}}^{2}=
\left(
  \begin{array}{cc}
    aa^{\dagger}&   0          \\
    0           & a^{\dagger}a 
  \end{array}
\right)=
\left(
  \begin{array}{cc}
    N+1& 0  \\
    0  & N
  \end{array}
\right)
\end{equation}
with the number operator $N$ we have 
\begin{equation}
\label{eq:solution-one}
\mbox{e}^{-itgA_{1}}=
\left(
  \begin{array}{cc}
  \mbox{cos}\left(tg\sqrt{N+1}\right) & 
  -i\frac{\mbox{sin}\left(tg\sqrt{N+1}\right)}{\sqrt{N+1}}a             \\
  -ia^{\dagger}\frac{\mbox{sin}\left(tg\sqrt{N+1}\right)}{\sqrt{N+1}} & 
  \mbox{cos}\left(tg\sqrt{N}\right)
  \end{array}
\right).
\end{equation}
We obtained the explicit form of solution. However, this form is more or less 
well--known, see for example the second book in \cite{books}.

\vspace{5mm}
\par \noindent 
{\bf Two Atoms Case}\quad In this case $A$ in (\ref{eq:A}) is written as 
\begin{equation}
\label{eq:A-two}
A_{2}=
\left(
  \begin{array}{cccc}
    0 &          a & a &           0  \\
    a^{\dagger}& 0 & 0 &           a  \\
    a^{\dagger}& 0 & 0 &           a  \\
    0 & a^{\dagger}& a^{\dagger} & 0
  \end{array}
\right).
\end{equation}

Our method is to reduce the $4\times 4$--matrix $A_{2}$ in (\ref{eq:A-two}) to 
a $3\times 3$--matrix $B_{1}$ in the following to make our calculation 
easier. 
For that aim we prepare the following matrix
\[
T=
\left(
  \begin{array}{cccc}
    0 &   1                & 0                  & 0   \\
    \frac{1}{\sqrt{2}} & 0 & \frac{1}{\sqrt{2}} & 0   \\
   -\frac{1}{\sqrt{2}} & 0 & \frac{1}{\sqrt{2}} & 0   \\
    0 &   0                &   0                  & 1
  \end{array}
\right),
\]
then it is easy to see 
\begin{equation}
\label{eq:decomposition-two}
T^{\dagger}A_{2}T=
\left(
  \begin{array}{cccc}
    0  &                     &                     &            \\
       & 0                   & \sqrt{2}a           & 0          \\
       & \sqrt{2}a^{\dagger} & 0                   & \sqrt{2}a  \\
       & 0                   & \sqrt{2}a^{\dagger} & 0
  \end{array}
\right)\equiv 
\left(
  \begin{array}{cc}
     0 &       \\
       & B_{1} 
  \end{array}
\right)
\end{equation}
where 
$
B_{1}=J_{+}\otimes a + J_{-}\otimes a^{\dagger}
$
and $\left\{J_{+},J_{-}\right\}$ are just generators of (spin one) 
irreducible representation of (\ref{eq:sigmas}). We note that this means 
a well--known decomposition of spin 
$\frac{1}{2}\otimes \frac{1}{2}=0\oplus 1$. 

Therefore to calculate $\mbox{e}^{-itgA_{2}}$ we have only to do 
$\mbox{e}^{-itgB_{1}}$. 
Noting the relation 
\[
{B_{1}}^{3}=
\left(
  \begin{array}{ccc}
    2(2N+3) &         &          \\
            & 2(2N+1) &          \\
            &         & 2(2N-1)
  \end{array}
\right)B\equiv DB_{1},
\]
we obtain 
\begin{equation}
\label{eq:solution-two-more(reduced)}
\mbox{e}^{-itgB_{1}}=
\left(
  \begin{array}{ccc}
   1+\frac{2N+2}{2N+3}f(N+1) & -ih(N+1)a & \frac{2}{2N+3}f(N+1)a^{2}   \\
   -ia^{\dagger}h(N+1) & 1+2f(N) & -ih(N)a                               \\
   (a^{\dagger})^{2}\frac{2}{2N+3}f(N+1) & -ia^{\dagger}h(N) & 
   1+\frac{2N}{2N-1}f(N-1)
  \end{array}
\right)
\end{equation}
where 
\[
f(N)=\frac{-1+\mbox{cos}\left(tg\sqrt{2(2N+1)}\right)}{2},\quad 
h(N)=\frac{\mbox{sin}\left(tg\sqrt{2(2N+1)}\right)}{\sqrt{2N+1}}.
\]

\vspace{5mm}\par \noindent 
{\bf Three Atoms Case}\quad In this case $A$ in (\ref{eq:A}) is written as 
\begin{equation}
\label{eq:A-three}
A_{3}=
\left(
  \begin{array}{cccccccc}
    0 &          a & a &           0  & a & 0 & 0 & 0          \\
    a^{\dagger}& 0 & 0 &           a  & 0 & a & 0 & 0          \\
    a^{\dagger}& 0 & 0 &           a  & 0 & 0 & a & 0          \\
    0 & a^{\dagger}& a^{\dagger} & 0  & 0 & 0 & 0 & a          \\
    a^{\dagger}& 0 & 0  &  0          & 0 & a & a & 0          \\
    0 & a^{\dagger}& 0  & 0   & a^{\dagger} &  0 & 0 & a       \\
    0 & 0 & a^{\dagger} & 0  & a^{\dagger} &  0 & 0 & a        \\
    0 & 0 & 0 & a^{\dagger} & 0 & a^{\dagger} & a^{\dagger} & 0    
  \end{array}
\right).
\end{equation}

We would like to look for the explicit form of solution like 
(\ref{eq:solution-one}) or (\ref{eq:solution-two-more(reduced)}). 
If we set 
\[
T=
\left(
  \begin{array}{cccccccc}
    0 & 0 & 0 & 0 & 1 & 0 & 0 & 0 \\
    \frac{1}{\sqrt{2}} & 0 & \frac{1}{\sqrt{6}} & 0 & 0 & 
    \frac{1}{\sqrt{3}} & 0 & 0 \\
    \frac{-1}{\sqrt{2}} & 0 & \frac{1}{\sqrt{6}} & 0 & 0 & 
    \frac{1}{\sqrt{3}} & 0 & 0 \\ 
    0 & 0 & 0 & \frac{\sqrt{2}}{\sqrt{3}} & 0 & 0 & \frac{1}{\sqrt{3}} & 0 \\
    0 & 0 & \frac{-\sqrt{2}}{\sqrt{3}} & 0 & 0 & \frac{1}{\sqrt{3}} & 0 & 0 \\
    0 & \frac{1}{\sqrt{2}} & 0 & \frac{-1}{\sqrt{6}} & 0 & 0 & 
    \frac{1}{\sqrt{3}} & 0 \\
    0 & \frac{-1}{\sqrt{2}} & 0 & \frac{-1}{\sqrt{6}} & 0 & 0 & 
    \frac{1}{\sqrt{3}} & 0 \\
    0 & 0 & 0 & 0 & 0 & 0 & 0 & 1
  \end{array}
\right),
\]
then it is not difficult to see 
\begin{equation}
\label{eq:decomposition-three}
T^{\dagger}A_{3}T=
\left(
  \begin{array}{cccccccc}
     0 & a &   &   &   &   &   &                       \\
    a^{\dagger}& 0 &   &   &    &   &   &              \\
       &   & 0 &  a &   &   &    &                     \\
       &   & a^{\dagger} & 0 &   &   &   &             \\
       &   &   &   & 0 & \sqrt{3}a & 0 & 0             \\
       &   &   &   & \sqrt{3}a^{\dagger} & 0 & 2a & 0  \\
       &   &   &   & 0 & 2a^{\dagger} & 0 & \sqrt{3}a  \\   
       &   &   &   & 0 & 0 & \sqrt{3}a^{\dagger} & 0
  \end{array}
\right)\equiv 
\left(
  \begin{array}{ccc}
     B_{1/2} &       &        \\
           & B_{1/2} &        \\ 
           &       & B_{3/2}
  \end{array}
\right).
\end{equation}
This means a decomposition of spin $\frac{1}{2}\otimes \frac{1}{2}\otimes 
\frac{1}{2}=\frac{1}{2}\oplus \frac{1}{2}\oplus \frac{3}{2}$. 
Therefore we have only to calculate $\mbox{e}^{-itgB_{3/2}}$, which is however 
not easy. The result is 
\begin{eqnarray}
\label{eq:solution-three-more(reduced)}
&&\mbox{e}^{-itgB_{3/2}}    \nonumber \\
=&&
\left(
  \begin{array}{cccc}
    f_{2}(N+2) & -\sqrt{3}iF_{1}(N+2)a & 2\sqrt{3}h_{1}(N+2)a^{2} & 
    -6iH_{0}(N+2)a^{3}   \\
    -\sqrt{3}iF_{1}(N+1)a^{\dagger} & f_{1}(N+1) & -2iH_{1}(N+1)a & 
    2\sqrt{3}h_{1}(N+1)a^{2} \\
    2\sqrt{3}h_{1}(N)(a^{\dagger})^{2} & -2iH_{1}(N)a^{\dagger} & 
    f_{0}(N) & -\sqrt{3}iF_{0}(N)a  \\
    -6iH_{0}(N-1)(a^{\dagger})^{3} & 2\sqrt{3}h_{1}(N-1)(a^{\dagger})^{2} & 
    -\sqrt{3}iF_{0}(N-1)a^{\dagger} & f_{-1}(N-1)   
  \end{array}
\right)      \nonumber \\ 
&{}& 
\end{eqnarray}
where 
\begin{eqnarray}
f_{2}(N)&=&\left\{v_{+}(N)\mbox{cos}(tg\sqrt{\lambda_{+}(N)})-
v_{-}(N)\mbox{cos}(tg\sqrt{\lambda_{-}(N)})\right\}/(2\sqrt{d(N)}), 
\nonumber \\
f_{1}(N)&=&\left\{w_{+}(N)\mbox{cos}(tg\sqrt{\lambda_{+}(N)})-
w_{-}(N)\mbox{cos}(tg\sqrt{\lambda_{-}(N)})\right\}/(2\sqrt{d(N)}), 
\nonumber \\
f_{0}(N)&=&\left\{v_{+}(N)\mbox{cos}(tg\sqrt{\lambda_{-}(N)})-
v_{-}(N)\mbox{cos}(tg\sqrt{\lambda_{+}(N)})\right\}/(2\sqrt{d(N)}), 
\nonumber \\
f_{-1}(N)&=&\left\{w_{+}(N)\mbox{cos}(tg\sqrt{\lambda_{-}(N)})-
w_{-}(N)\mbox{cos}(tg\sqrt{\lambda_{+}(N)})\right\}/(2\sqrt{d(N)}), 
\nonumber \\
h_{1}(N)&=&\left\{\mbox{cos}(tg\sqrt{\lambda_{+}(N)})-
\mbox{cos}(tg\sqrt{\lambda_{-}(N)})\right\}/(2\sqrt{d(N)}),
\nonumber \\
F_{1}(N)&=&\left\{\frac{w_{+}(N)}{\sqrt{\lambda_{+}(N)}}
\mbox{sin}(tg\sqrt{\lambda_{+}(N)})-
\frac{w_{-}(N)}{\sqrt{\lambda_{-}(N)}}
\mbox{sin}(tg\sqrt{\lambda_{-}(N)})\right\}/(2\sqrt{d(N)}), 
\nonumber \\
F_{0}(N)&=&\left\{\frac{v_{+}(N)}{\sqrt{\lambda_{-}(N)}}
\mbox{sin}(tg\sqrt{\lambda_{-}(N)})-
\frac{v_{-}(N)}{\sqrt{\lambda_{+}(N)}}
\mbox{sin}(tg\sqrt{\lambda_{+}(N)})\right\}/(2\sqrt{d(N)}), 
\nonumber \\
H_{1}(N)&=&\left\{\sqrt{\lambda_{+}(N)}
\mbox{sin}(tg\sqrt{\lambda_{+}(N)})-
\sqrt{\lambda_{-}(N)}
\mbox{sin}(tg\sqrt{\lambda_{-}(N)})\right\}/(2\sqrt{d(N)}),
\nonumber \\
H_{0}(N)&=&\left\{\frac{1}{\sqrt{\lambda_{+}(N)}}
\mbox{sin}(tg\sqrt{\lambda_{+}(N)})-
\frac{1}{\sqrt{\lambda_{-}(N)}}
\mbox{sin}(tg\sqrt{\lambda_{-}(N)})\right\}/(2\sqrt{d(N)})
\nonumber 
\end{eqnarray}
and 
\begin{eqnarray}
&&\lambda_{\pm}(N)=5N\pm \sqrt{d(N)},\ 
v_{\pm}(N)=-2N-3\pm \sqrt{d(N)},\ 
w_{\pm}(N)=2N-3\pm \sqrt{d(N)},     \nonumber \\
&&d(N)=16N^{2}+9. \nonumber 
\end{eqnarray}

This form is very complicated. We note that to calculate 
$\mbox{e}^{-itgB_{3/2}}$ we used Mathematica to the fullest.

\section{Quantum Diagonalization Method}

First of all we explain the method which we call a {\bf Q}uantum 
{\bf D}iagonalization {\bf M}ethod (QDM). To calculate $\mbox{e}^{-itgA}$ 
for $A$ in (\ref{eq:A}) 
\[
A=S_{+}\otimes a + S_{-}\otimes a^{\dagger}
\]
we would like to diagonalize it like $A=UD_{A}U^{\dagger}$ ($D_{A}$ is a 
diagonal matrix) if possible. This is a well--known classsical procedure. 
However, in our case it is impossible because we cannot determine the 
eigenvalues by making use of its characteristic equation $f(\lambda)=
\mbox{det}(\lambda {\bf 1}-A)$. In the quantum case there is no meaning on 
determinant function. For example, which is correct 
\[
f(\lambda)=
\left|
  \begin{array}{cc}
    \lambda      &   -a      \\
    -a^{\dagger} & \lambda 
  \end{array}
\right|
\stackrel{\mbox{?}}{=}
\left\{
\begin{array}{ll}
  \lambda^{2}-aa^{\dagger}  \\
  \lambda^{2}-a^{\dagger}a
\end{array}
\right.
\]
Therefore we have no general method to make $A$ diagonal. However, we have 
a very skillful method for $A$ whose procedure goes like 
\begin{center}
{\bf Classicalization} $\longrightarrow$ {\bf Quantization} $\longrightarrow$ 
{\bf Classicalization}.
\end{center}

The (quantum) matrix $A$ above can be decomposed as 
\[
T^{\dagger}AT=
{\sum}_{\oplus}B_{j}\quad (\mbox{a direct sum of quantum matrices of spin}\ j) 
\]
by an orthogonal matrix $T$ \footnote{To find $T$ in the general case is not 
easy} like in the preceeding section, where $B_{j}$ is given by 
\begin{eqnarray}
\label{eq:spin-j qmatrix}
&&B_{j}=     \nonumber \\
&&
\left(
  \begin{array}{ccccccc}
    0 & \sqrt{(J-1)1}a &    &    &     &     &    \\
    \sqrt{(J-1)1}a^{\dagger} & 0 & \sqrt{(J-2)2}a &   &   &   &    \\
      & \sqrt{(J-2)2}a^{\dagger} & 0 & \sqrt{(J-3)3}a &   &   &    \\
      &    & \cdot & \cdot & \cdot &    &           \\
      &    &    &  \cdot & \cdot & \cdot &          \\
      &    &    &    & \sqrt{2(J-2)}a^{\dagger} & 0 & \sqrt{1(J-1)}a \\
      &    &    &    & 0 & \sqrt{1(J-1)}a^{\dagger} & 0
\end{array}
\right)      \nonumber \\
&&=\left(\sqrt{(J-k)k}a\delta_{k,i-1}+
\sqrt{(J-i)i}a^{\dagger}\delta_{k-1,i}\right)
\end{eqnarray}
with $J=2j+1$. In the following we set $B=B_{j}$ for simplicity. 

\par \vspace{5mm} \noindent 
{\bf (i)\ Classicalization}\quad We replace $a\rightarrow z$ and 
$a^{\dagger}\rightarrow {\bar z}$ in $B$ and set 
\begin{eqnarray}
&&C=     \nonumber \\
&&
\left(
  \begin{array}{ccccccc}
    0 & \sqrt{(J-1)1}z &    &    &     &     &                   \\
    \sqrt{(J-1)1}\bar{z} & 0 & \sqrt{(J-2)2}z &   &   &   &      \\
      & \sqrt{(J-2)2}\bar{z} & 0 & \sqrt{(J-3)3}z &   &   &      \\
      &    & \cdot & \cdot & \cdot &    &                        \\
      &    &    &  \cdot & \cdot & \cdot &                       \\
      &    &    &    & \sqrt{2(J-2)}\bar{z} & 0 & \sqrt{1(J-1)}z \\
      &    &    &    & 0 & \sqrt{1(J-1)}\bar{z} & 0
\end{array}
\right)      \nonumber \\
&&=\left(\sqrt{(J-k)k}z\delta_{k,i-1}+
\sqrt{(J-i)i}\bar{z}\delta_{k-1,i}\right). 
\end{eqnarray}
We must diagonalize $C$. The eigenvalues are 
\[
\left\{(J-1)|z|, (J-3)|z|, 
\cdots, (J-2i+1)|z|, \cdots, -(J-3)|z|, -(J-1)|z|\right\}
\]
and corresponding orthonormal eigenvectors are 
\[
\ket{(J-2i+1)|z|}
=\left(x_{ki}\frac{{\bar{z}}^{k-1}}{|z|^{k-1}}\right)_{k=1\sim J}
\quad \mbox{for}\quad i=1\sim J
\]
where $x_{ki}$ are defined as 
\[
x_{ki}=\frac{y_{ki}}{\sqrt{\sum_{l=1}^{J}y_{li}^{2}}}
\]
with $y_{ki}$ ($k=1\sim J$) defined by the recursion relation 
\[
y_{1i}=1,\quad
\sqrt{(J-k+1)(k-1)}y_{k-1,i}+\sqrt{(J-k)k}y_{k+1,i}=(J-2i+1)y_{ki}.
\]
For example, 
\[
y_{1i}=1,\quad 
y_{2i}=\frac{J-2i+1}{\sqrt{(J-1)1}},\quad 
y_{3i}=\frac{(J-2i+1)^{2}-(J-1)1}{\sqrt{(J-1)(J-2)1\cdot 2}},\quad \mbox{etc}.
\]
We note that the matrix $X=(x_{ki})$ is an (real) orthonormal one, namely 
$X^{T}X=XX^{T}={\bf 1}_{J}$. For example, when $j=1$ ($J=3$) it is easy to 
show 
\[
X=
\left(
  \begin{array}{ccc}
    \frac{1}{2} & \frac{1}{\sqrt{2}} & \frac{1}{2}  \\
    \frac{1}{\sqrt{2}} & 0 & -\frac{1}{\sqrt{2}}    \\
    \frac{1}{2} & -\frac{1}{\sqrt{2}} & \frac{1}{2} 
\end{array}
\right).
\]

If we set 
\begin{equation}
\label{eq:classical-unitary}
W=\left(x_{ki}\frac{{\bar{z}}^{k-1}}{|z|^{k-1}}\right)
\end{equation}
then $W$ is a unitary matrix ($W^{\dagger}W=WW^{\dagger}={\bf 1}_{J}$) and 
$C$ is diagonalized by $W$ as 
\begin{equation}
\label{eq:classical-diagonal}
C=WD_{C}W^{\dagger}
\end{equation}
where $D_{C}$ is a diagonal matrix consisting of the eigenvalues $\{
(J-2i+1)|z|\ |\ i=1\sim J\}$.

We note that the unitary matrix $W$ is not defined at $z=0$, see 
(\ref{eq:classical-unitary}).

\par \vspace{5mm} \noindent 
{\bf (ii)\ Quantization}\quad Next we consider a quantization of $W$ : namely 
we want to find a (quantum) unitary matrix $U_{1}$ arising from $W$ above. 
After some trial and errors we set 
\begin{equation}
U_{1}=
\left(\frac{x_{ki}}{\sqrt{N(N-1)\cdots (N-k+2)}}(a^{\dagger})^{k-1}\right), 
\end{equation}
then it is not difficult to check 
\[
U_{1}U_{1}^{\dagger}=U_{1}^{\dagger}U_{1}={\bf 1}_{J} 
\]
on the representation space $\calh\oplus \calh_{1}\oplus \cdots \oplus 
\calh_{J-1}$ where $\calh_{k}=\mbox{Vect}_{\fukuso}\{\ket{k},\ket{k+1},\cdots 
\ \}$ is the subspace of the Fock space $\calh\equiv \calh_{0}$ generated by 
$\{a,a^{\dagger}, N\}$. 
We note that $U_{1}$ is not defined on the whole space \footnote{$U_{1}$ is 
a partial isometry on $\calh\oplus \calh\oplus \cdots \oplus \calh$ in the 
mathematical terminology}.

A comment is in order.\ Noting 
\begin{eqnarray}
&&(a^{\dagger})^{l}a^{l}=N(N-1)\cdots (N-l+1),\quad 
a^{l}(a^{\dagger})^{l}=(N+l)(N+l-1)\cdots (N+1),  \nonumber \\
&&aN=(N+1)a,\quad a^{\dagger}N=(N-1)a^{\dagger}     \nonumber 
\end{eqnarray}
for $l\geq 1$ we have 
\begin{eqnarray}
\frac{1}{\sqrt{(a^{\dagger})^{l}a^{l}}}(a^{\dagger})^{l}
&=&
\frac{1}{\sqrt{N(N-1)\cdots (N-l+1)}}(a^{\dagger})^{l}      \nonumber \\
&=&
(a^{\dagger})^{l}\frac{1}{\sqrt{(N+l)(N+l-1)\cdots (N+1)}}
=
(a^{\dagger})^{l}\frac{1}{\sqrt{a^{l}(a^{\dagger})^{l}}}
\end{eqnarray}
and 
\begin{equation}
(a^{\dagger})^{l}\frac{1}{a^{l}(a^{\dagger})^{l}}a^{l}
=(a^{\dagger})^{l}\frac{1}{(N+1)\cdots (N+l-1)(N+l)}a^{l}
={\bf 1}\quad \mbox{on}\quad \calh_{l}.
\end{equation}

For example, when $j=1$ ($J=3$) we have 
\[
U_{1}=
\left(
\begin{array}{ccc}
 \frac{1}{2} & \frac{1}{\sqrt{2}} & \frac{1}{2}      \\
 \frac{1}{\sqrt{2}\sqrt{N}}a^{\dagger} & 0 & 
 -\frac{1}{\sqrt{2}\sqrt{N}}a^{\dagger}              \\
 \frac{1}{2\sqrt{N(N-1)}}(a^{\dagger})^{2} & 
 -\frac{1}{\sqrt{2}\sqrt{N(N-1)}}(a^{\dagger})^{2} & 
 \frac{1}{2\sqrt{N(N-1)}}(a^{\dagger})^{2} 
\end{array}
\right)
\]

\par \vspace{5mm} \noindent 
{\bf (iii)\ Classicalization}\quad Here we consider a diagonalization of $B$ 
by $U_{1}$ above. Some calculation leads 
\begin{equation}
U_{1}^{\dagger}BU_{1}=R=(r_{ki})
\end{equation}
where 
\[
r_{ki}=\sum_{l=2}^{J}\sqrt{(J-l+1)(l-1)}(y_{l-1,k}y_{li}+y_{l,k}y_{l-1,i})
\sqrt{N+l-1}=r_{ik}.
\]
We see that the matrix $R$ is hermitian and its entries consist of some 
functions of the number operator $N$, so $R$ is a kind of classical matrix. 
Therefore we can make $R$ diagonal \footnote{To obtain $U_{2}$ explicitly is 
not easy or almost impossible} like 
\begin{equation}
R=U_{2}D_{R}U_{2}^{\dagger}.
\end{equation}
For example, when $j=1$ ($J=3$) we have 
\[
R=
\left(
\begin{array}{ccc}
 \sqrt{N+1}+\sqrt{N+2} & -\frac{\sqrt{N+2}-\sqrt{N+1}}{\sqrt{2}} & 0    \\
 -\frac{\sqrt{N+2}-\sqrt{N+1}}{\sqrt{2}} & 0 & 
 \frac{\sqrt{N+2}-\sqrt{N+1}}{\sqrt{2}}                                 \\
 0 & \frac{\sqrt{N+2}-\sqrt{N+1}}{\sqrt{2}} & -(\sqrt{N+1}+\sqrt{N+2})
\end{array}
\right)
\]
and
\begin{eqnarray}
U_{2}&=&
\left(
\begin{array}{ccc}
 -\frac{\sqrt{2(2N+3)}+\sqrt{N+2}+\sqrt{N+1}}{2\sqrt{2(2N+3)}} & 
 \frac{\sqrt{N+2}-\sqrt{N+1}}{\sqrt{2}\sqrt{2(2N+3)}} & 
 -\frac{\sqrt{2(2N+3)}-\sqrt{N+2}-\sqrt{N+1}}{2\sqrt{2(2N+3)}}     \\
 \frac{\sqrt{N+2}-\sqrt{N+1}}{\sqrt{2}\sqrt{2(2N+3)}} & 
 \frac{\sqrt{N+2}+\sqrt{N+1}}{\sqrt{2(2N+3)}} &
 -\frac{\sqrt{N+2}-\sqrt{N+1}}{\sqrt{2}\sqrt{2(2N+3)}}             \\ 
 \frac{\sqrt{2(2N+3)}-\sqrt{N+2}-\sqrt{N+1}}{2\sqrt{2(2N+3)}} &
 \frac{\sqrt{N+2}-\sqrt{N+1}}{\sqrt{2}\sqrt{2(2N+3)}} & 
 \frac{\sqrt{2(2N+3)}+\sqrt{N+2}+\sqrt{N+1}}{2\sqrt{2(2N+3)}}
\end{array}
\right)         \nonumber \\
D&=&
\left(
\begin{array}{ccc}
 \sqrt{2(2N+3)} &   &                  \\
                & 0 &                  \\
                &   & -\sqrt{2(2N+3)}
\end{array}
\right).        \nonumber 
\end{eqnarray}

As a result we finally obtain 
\begin{equation}
\label{eq:quantum diagonalization}
B=\left(U_{1}U_{2}\right)D_{R}{\left(U_{1}U_{2}\right)}^{\dagger}
\equiv UDU^{\dagger}.
\end{equation}

\par \noindent
This is just the diagonal form of $B$ that we are looking for. We note that 
all entries of $U=U_{1}U_{2}$ consist of $N$ and $a^{\dagger}$ 
(not contain $a$). From this we have 
\begin{equation}
\mbox{e}^{-itgB}=U\mbox{e}^{-itgD}U^{\dagger}.
\end{equation}

This is a kind of ``normal ordered" diagonal expression of the evolution 
operator. See Appendix for another diagonal expression. 
In the following we give an explicit expression in the case of one, 
two and three atoms. 

\vspace{5mm}
Let us list the results. 

\vspace{3mm}
\par \noindent 
{\bf One Atom Case}\quad For $A_{1}$ in (\ref{eq:A-one}) we have 
\begin{equation}
\label{eq:diagonalization-1}
A_{1}=UDU^{\dagger}
\end{equation}
where 
\begin{equation}
\label{eq:diagonalization-1-com}
U=\frac{1}{\sqrt{2}}
\left(
\begin{array}{cc}
 1                             & 1                              \\
 \frac{1}{\sqrt{N}}a^{\dagger} & -\frac{1}{\sqrt{N}}a^{\dagger} 
\end{array}
\right),\quad
D=
\left(
\begin{array}{cc}
 \sqrt{N+1} &              \\
            & -\sqrt{N+1}
\end{array}
\right).
\end{equation}
Then it is easy to see 
\begin{equation}
\label{eq:reproduction-one}
\mbox{e}^{-itgA_{1}}=U\mbox{e}^{-itgD}U^{\dagger}=
\mbox{the right hand side of (\ref{eq:solution-one})}.
\end{equation}

\vspace{5mm}
\par \noindent 
{\bf Two Atoms Case}\quad For $B_{1}$ in (\ref{eq:decomposition-two}) we have 

\begin{equation}
\label{eq:diagonalization-2}
B_{1}=UDU^{\dagger}
\end{equation}
where 
\begin{equation}
\label{eq:diagonalization-2-com-U}
U=
\left(
\begin{array}{ccc}
 -\frac{\sqrt{N+1}}{\sqrt{2(2N+3)}} & 
 \frac{\sqrt{2}\sqrt{N+2}}{\sqrt{2(2N+3)}}  &     
 \frac{\sqrt{N+1}}{\sqrt{2(2N+3)}}                           \\
 -\frac{1}{\sqrt{2}\sqrt{N}}a^{\dagger}  & 0 &
 -\frac{1}{\sqrt{2}\sqrt{N}}a^{\dagger}                      \\
 -\frac{1}{\sqrt{N-1}\sqrt{2(2N-1)}}(a^{\dagger})^{2} &
 -\frac{\sqrt{2}}{\sqrt{N}\sqrt{2(2N-1)}}(a^{\dagger})^{2} &
 \frac{1}{\sqrt{N-1}\sqrt{2(2N-1)}}(a^{\dagger})^{2} 
\end{array}
\right)
\end{equation}
and 
\begin{equation}
\label{eq:diagonalization-2-com-D}
D=
\left(
\begin{array}{ccc}
 \sqrt{2(2N+3)} &   &                  \\
                & 0 &                  \\
                &   & -\sqrt{2(2N+3)}
\end{array}
\right).
\end{equation}
Then it is not difficult to see 
\begin{equation}
\label{eq:reproduction-two}
\mbox{e}^{-itgB_{1}}=U\mbox{e}^{-itgD}U^{\dagger}=
\mbox{the right hand side of (\ref{eq:solution-two-more(reduced)})}.
\end{equation}

\vspace{5mm}
\par \noindent 
{\bf Three Atoms Case}\quad For $B_{3/2}$ in (\ref{eq:decomposition-three}) 
we have \footnote{The calculation in this case is interesting and hard, so 
the full details will be given in \cite{KHi}}
\begin{equation}
\label{eq:diagonalization-3}
B_{3/2}=UDU^{\dagger}
\end{equation}
where 
\begin{equation}
\label{eq:diagonalization-3-com-U}
U=
\left(
\begin{array}{cccc}
 u_{11} & u_{12} & u_{13} & u_{14} \\
 u_{21} & u_{22} & u_{23} & u_{24} \\
 u_{31} & u_{32} & u_{33} & u_{34} \\ 
 u_{41} & u_{42} & u_{43} & u_{44} 
\end{array}
\right)
\end{equation}
where 
\begin{eqnarray}
u_{11}
&=&\frac{1}{4}\left\{\sqrt{2}(1-\beta\gamma)+\sqrt{6}(\beta+\gamma)\right\}
 \frac{1}{\sqrt{(1+\beta^{2})(1+\gamma^{2})}}, \nonumber \\
u_{12}
&=&\frac{1}{4}\left\{\sqrt{6}(1-\beta\gamma)-\sqrt{2}(\beta+\gamma)\right\}
 \frac{1}{\sqrt{(1+\beta^{2})(1+\gamma^{2})}}, \nonumber \\
u_{13}&=&u_{12},\quad  u_{14}=-u_{11},       \nonumber \\
u_{21}
&=&\frac{1}{4}\left\{\frac{\sqrt{6}}{\sqrt{N}}a^{\dagger}(1+\beta\gamma)+
 \frac{\sqrt{2}}{\sqrt{N}}a^{\dagger}(\beta-\gamma)\right\}
 \frac{1}{\sqrt{(1+\beta^{2})(1+\gamma^{2})}}, \nonumber \\
u_{22}
&=&\frac{1}{4}\left\{\frac{\sqrt{2}}{\sqrt{N}}a^{\dagger}(1+\beta\gamma)-
 \frac{\sqrt{6}}{\sqrt{N}}a^{\dagger}(\beta-\gamma)\right\}
 \frac{1}{\sqrt{(1+\beta^{2})(1+\gamma^{2})}}, \nonumber \\
u_{23}&=&-u_{22},\quad u_{24}=u_{21},  \nonumber \\
u_{31}
&=&\frac{1}{4}\left\{
 \frac{\sqrt{6}}{\sqrt{N(N-1)}}(a^{\dagger})^{2}(1-\beta\gamma)-
 \frac{\sqrt{2}}{\sqrt{N(N-1)}}(a^{\dagger})^{2}(\beta+\gamma)\right\}
 \frac{1}{\sqrt{(1+\beta^{2})(1+\gamma^{2})}}, \nonumber \\
u_{32}
&=-&\frac{1}{4}\left\{
 \frac{\sqrt{2}}{\sqrt{N(N-1)}}(a^{\dagger})^{2}(1-\beta\gamma)+
 \frac{\sqrt{6}}{\sqrt{N(N-1)}}(a^{\dagger})^{2}(\beta+\gamma)\right\}
 \frac{1}{\sqrt{(1+\beta^{2})(1+\gamma^{2})}}, \nonumber \\
u_{33}&=&u_{32},\quad u_{34}=-u_{31},  \nonumber \\
u_{41}
&=&\frac{1}{4}\left\{
 \frac{\sqrt{2}}{\sqrt{N(N-1)(N-2)}}(a^{\dagger})^{3}(1+\beta\gamma)-
 \frac{\sqrt{6}}{\sqrt{N(N-1)(N-2)}}(a^{\dagger})^{3}(\beta-\gamma)\right\}
\nonumber \\
 &&\quad \times \frac{1}{\sqrt{(1+\beta^{2})(1+\gamma^{2})}}, \nonumber \\
u_{42}
&=&-\frac{1}{4}\left\{
 \frac{\sqrt{6}}{\sqrt{N(N-1)(N-2)}}(a^{\dagger})^{3}(1+\beta\gamma)+
 \frac{\sqrt{2}}{\sqrt{N(N-1)(N-2)}}(a^{\dagger})^{3}(\beta-\gamma)\right\}
\nonumber \\
 &&\quad \times \frac{1}{\sqrt{(1+\beta^{2})(1+\gamma^{2})}}, \nonumber \\
u_{43}&=&-u_{42},\quad u_{44}=u_{41}
\end{eqnarray}
and 
\begin{eqnarray}
\label{eq:relations-1}
&&\beta=\frac{\mu-\nu-(x-y)}{2b},\quad \gamma=\frac{\mu+\nu-(x+y)}{2c},
\nonumber \\
&&x=3\sqrt{N+1}+6\sqrt{N+2}+3\sqrt{N+3},\quad 
y=3\sqrt{N+1}-2\sqrt{N+2}+3\sqrt{N+3},
\nonumber \\
&&b=2\sqrt{3}\left(\sqrt{N+1}-\sqrt{N+3}\right),\quad 
c=\sqrt{3}\left(\sqrt{N+1}-2\sqrt{N+2}+\sqrt{N+3}\right),
\end{eqnarray}
and 
\begin{eqnarray}
\label{eq:relations-2}
\mu&=&4\sqrt{5(N+2)+\sqrt{16(N+2)^{2}+9}}=4\sqrt{\lambda_{+}(N+2)},
\nonumber \\
\nu&=&4\sqrt{5(N+2)-\sqrt{16(N+2)^{2}+9}}=4\sqrt{\lambda_{-}(N+2)}
\end{eqnarray}
and 
\begin{eqnarray}
\label{eq:diagonalization-3-com-D}
D&=&\frac{1}{4}
\left(
\begin{array}{cccc}
 \mu &     &      &       \\
     & \nu &      &       \\
     &     & -\nu &       \\
     &     &      & -\mu 
\end{array}
\right)      \nonumber \\
&=&
\left(
\begin{array}{cccc}
  \sqrt{\lambda_{+}(N+2)} &     &      &      \\
     & \sqrt{\lambda_{-}(N+2)} &      &       \\
     &     & -\sqrt{\lambda_{-}(N+2)} &       \\
     &     &      & -\sqrt{\lambda_{+}(N+2)}
\end{array}
\right).
\end{eqnarray}
Then we obtain 
\begin{equation}
\label{eq:reproduction-three}
\mbox{e}^{-itgB_{3/2}}=U\mbox{e}^{-itgD}U^{\dagger}=
\mbox{the right hand side of (\ref{eq:solution-three-more(reduced)})}.
\end{equation}
However, the proof is not easy, see Appendix. 

Last we make one comment. We would like to perform a diagonalization 
for the case of more than three atoms, however it is not easy at the present. 
One of main difficulties is the step (iii). 
That is, to perform a diagonalization to the (classical hermite) matrix 
we must determine its eigenvalues by solving the characteristic equation 
and give orthonormal eigenvectors explicitly (not abstractly). 
The characteristic equation is in general algebraic one of degrees more than 
four, which is impossible to solve in an algebraic manner by the famous 
Galois theory. Even for algebraic equations of degrees three and four we must 
use the Cardano and Ferrarri formulas (see for example \cite{KF-special}) 
which make hard to determine all orthonormal eigenvectors explicitly.

\section{U(1) Ambiguity}

In this section we discuss a problem of $U(1)$ ambiguity in the quantum 
diagonalization method. 

The classical diagonalization $C=WD_{C}W^{\dagger}$ in 
(\ref{eq:classical-diagonal}) has the following $U(1)$ ``invariance", namely 
\[
C=WD_{C}W^{\dagger}=WU_{0}(U_{0}^{\dagger}D_{C}U_{0})U_{0}^{\dagger}W^{\dagger}
=(WU_{0})D_{C}(WU_{0})^{\dagger}
\]
where $U_{0}$ is a diagonal matrix defined by 
\[
U_{0}
=
\left(
\begin{array}{cccccc}
f_{1}(z,\bar{z}) &   &   &   &   &    \\
 & f_{2}(z,\bar{z}) &    &   &   &    \\
 &   & \cdot &   &   &                \\
 &   &   & \cdot &   &                \\
 &   &   &   & f_{J-1}(z,\bar{z}) &   \\
 &   &   &   &   & f_{J}(z,\bar{z})
\end{array}
\right)
\quad \in \quad U(J) 
\]
and $f_{j}(z,\bar{z})$ is an element in $U(1)$, $|f_{j}(z,\bar{z})|^{2}=1$. 
For example, $f_{j}(z,\bar{z})=\frac{z^{j-1}}{|z|^{j-1}}$. 
Namely, the diagonal matrix $D_{C}$ is invariant under the change of unitary 
matrix $W\longrightarrow WU_{0}$. 

However, this is not kept in the process of quantization. 
From (\ref{eq:quantum diagonalization}) 
\[
B=UDU^{\dagger}=(U\tilde{U}_{0})(\tilde{U}_{0}^{\dagger}D\tilde{U}_{0})
(U\tilde{U}_{0})^{\dagger}\equiv \tilde{U}\tilde{D}{\tilde{U}}^{\dagger}
\]
where $\tilde{U}_{0}$ is a quantum ``diagonal" matrix defined by 
\[
\tilde{U}_{0}
=
\left(
\begin{array}{cccccc}
f_{1}(a,a^{\dagger}) &   &   &   &   &    \\
 & f_{2}(a,a^{\dagger}) &    &   &   &    \\
 &   & \cdot &   &   &                    \\
 &   &   & \cdot &   &                    \\
 &   &   &   & f_{J-1}(a,a^{\dagger}) &   \\
 &   &   &   &   & f_{J}(a,a^{\dagger})
\end{array}
\right)
\]
and $f_{j}(a,a^{\dagger})$ is an element satisfying 
$f_{j}(a,a^{\dagger})^{\dagger}f_{j}(a,a^{\dagger})=f_{j}(a,a^{\dagger})
f_{j}(a,a^{\dagger})^{\dagger}={\bf 1}$. 
We note that $\tilde{U}_{0}$ is not defined on the whole space, which 
changes a domain and a range of $\tilde{U}$. 

Here we restrict each $f_{j}(a,a^{\dagger})$ to one satisfying a relation 
\[
f_{j}(a,a^{\dagger})^{\dagger}Nf_{j}(a,a^{\dagger})=g_{j}(N)
\]
for some function $g_{j}$. For example, 
$f_{j}(a,a^{\dagger})=\frac{1}{\sqrt{(N+j-1)(N+j-2)\cdots(N+1)}}a^{j-1}$. 
In this case 
\[
D\ne \tilde{D}=\tilde{U}_{0}^{\dagger}D\tilde{U}_{0}
\]
because $f_{j}(a,a^{\dagger})$ and the number operator $N$ don't 
commute in general. 

Let us show this with an example. For $B_{1}$ in the two atoms case we 
consider a very simple case 
\[
\tilde{U}_{0}
=
\left(
\begin{array}{ccc}
\frac{1}{\sqrt{N+1}}a &  &   \\
& \frac{1}{\sqrt{N+1}}a  &   \\
&   & \frac{1}{\sqrt{N+1}}a
\end{array}
\right), 
\]
then it is easy to see 
\[
B_{1}={\tilde U}{\tilde D}{\tilde U}^{\dagger}
\]
where 
\[
{\tilde U}=
\left(
\begin{array}{ccc}
 -\frac{1}{\sqrt{2(2N+3)}}a & 
 \frac{\sqrt{2}\sqrt{N+2}}{\sqrt{N+1}\sqrt{2(2N+3)}}a &
 \frac{1}{\sqrt{2(2N+3)}}a   \\
 -\frac{1}{\sqrt{2}} & 0 & -\frac{1}{\sqrt{2}} \\
 -\frac{1}{\sqrt{2(2N-1)}}a^{\dagger} & 
 -\frac{\sqrt{2}\sqrt{N-1}}{\sqrt{N}\sqrt{2(2N-1)}}a^{\dagger} &
 \frac{1}{\sqrt{2(2N-1)}}a^{\dagger}
\end{array}
\right)
\]
and 
\[
{\tilde D}=
\left(
\begin{array}{ccc}
 \sqrt{2(2N+1)} &    &                 \\
                & 0  &                 \\
                &    & -\sqrt{2(2N+1)}
\end{array}
\right). 
\]
Compare this with (\ref{eq:diagonalization-2}). Here we note that 
\[
{\tilde U}^{\dagger}{\tilde U}={\bf 1}_{\calh\oplus \calh_{1}\oplus \calh}
\quad \mbox{and} \quad 
{\tilde U}{\tilde U}^{\dagger}={\bf 1}_{\calh\oplus \calh\oplus \calh_{1}}.
\]
In the two expressions 
$B_{1}=UDU^{\dagger}={\tilde U}{\tilde D}{\tilde U}^{\dagger}$, each domain 
and range of $U$ and ${\tilde U}$ is different. 

\vspace{3mm}
The diagonal part $D$ of $B$ changes according to $\tilde{U}_{0}$, which is 
unavoidable due to the noncommutativity of operators in quantum physics. 
We call this phenomenon a $U(1)$ ambiguity.

\section{Discussion}

We introduced the quantum diagonalization method and applied it to the 
(quantum) matrix $B$ in (\ref{eq:spin-j qmatrix}) (or $A$ in (\ref{eq:A})) 
emerging from the Tavis--Cummings model and (re)obtained the explicit form of 
evolution operator for the one, two and three atoms case. 
To get the general case is not easy because of some technical reasons 
(numerical techniques are of course applicable). 

Therefore, 
there are many applications to quantum optics or mathematical physics, see 
for example \cite{papers-1}. 
We can also apply the result to a quantum computation based on atoms of 
laser--cooled and trapped linearly in a cavity, see \cite{FHKW}. 

We also make a comment on an application to a noncommutative differential 
geometry. From (\ref{eq:quantum diagonalization}) we have a (quantum) 
unitary matrix $U$ which gives the Maurer--Cartan forms 
\[
L_{U}\equiv U^{-1}\hat{d}U,\quad R_{U}\equiv \hat{d}U{U^{-1}}.
\]
where $\hat{d}$ is some differential with respect to $a$ and $a^{\dagger}$. 
These are fundamental objects in noncommutative chiral models. For the case 
of one, two and three atoms we can calculate the Maurer-Cartan forms exactly. 
Such a study is however beyond our scope of this paper. We expect that some 
researchers will develop the subject. 

We conclude this paper by making a comment. The Tavis--Cummings model 
is based on (only) two energy levels of atoms. However, an atom has in general 
infinitely many energy levels, so it is natural to use this possibility. 
We are also studying a quantum computation based on multi--level systems of 
atoms (a qudit theory) \cite{qudit-papers}. Therefore we would like to extend 
the Tavis--Cummings model based on two--levels to a model based on 
multi--levels. This is a very challenging task.

\vspace{5mm}
\noindent
{\it Acknowledgment.}
We wish to thank Shin'ichi Nojiri for his helpful comments and suggestions. 

\par \vspace{10mm}
\begin{center}
 \begin{Large}
   {\bf Appendix}
 \end{Large}
\end{center}

\par \vspace{5mm} \noindent
{\bf \ \ Proof of (\ref{eq:reproduction-three})}

Here we show (\ref{eq:reproduction-three}). The (1,1)--entry of 
$U\mbox{e}^{-itgD}U^{\dagger}$ is 
\[
\mbox{e}^{-itg\mu/4}u_{11}^{2}+\mbox{e}^{-itg\nu/4}u_{12}^{2}+
\mbox{e}^{itg\nu/4}u_{13}^{2}+\mbox{e}^{itg\mu/4}u_{14}^{2}
=
2u_{11}^{2}\mbox{cos}\left(tg\frac{\mu}{4}\right)+
2u_{12}^{2}\mbox{cos}\left(tg\frac{\nu}{4}\right).
\]
because $u_{13}^{2}=u_{12}^{2}$ and $u_{14}^{2}=u_{11}^{2}$. 
Let us calculate both $u_{11}^{2}$ and $u_{12}^{2}$.
\begin{eqnarray}
2u_{11}^{2}
&=&
\frac{\left\{\sqrt{2}(1-\beta\gamma)+\sqrt{6}(\beta+\gamma)\right\}^{2}}
{8(1+\beta^{2})(1+\gamma^{2})}     \nonumber \\
&=&
\frac{1}{8}\left\{2+
4\frac{\sqrt{3}\left(\frac{1}{\beta}-\beta+\frac{1}{\gamma}-\gamma\right)+
\frac{\beta}{\gamma}+\frac{\gamma}{\beta}+2}
{\left(\beta+\frac{1}{\beta}\right)\left(\gamma+\frac{1}{\gamma}\right)}
\right\}       \nonumber \\
&=&
\frac{1}{8}\left\{2+
4\frac{\sqrt{3}\{c(x-y)+b(x+y)\}+\frac{\mu^{2}-\nu^{2}}{2}-
\frac{x^{2}-y^{2}}{2}+2bc}{\mu^{2}-\nu^{2}}\right\}
\nonumber \\
&=&
\frac{1}{8}\left\{
2+\frac{-8N-28+2\sqrt{16(N+2)^{2}+9}}{\sqrt{16(N+2)^{2}+9}}\right\}
\nonumber \\
&=&
\frac{-2N-7+\sqrt{16(N+2)^{2}+9}}{2\sqrt{16(N+2)^{2}+9}}
=\frac{v_{+}(N+2)}{2\sqrt{d(N+2)}}    \nonumber 
\end{eqnarray}
where we have used the relations 
\begin{eqnarray}
&&\beta+\frac{1}{\beta}=\frac{\mu-\nu}{b},\quad 
\frac{1}{\beta}-\beta=\frac{x-y}{b},\quad 
\gamma+\frac{1}{\gamma}=\frac{\mu+\nu}{c},\quad 
\frac{1}{\gamma}-\gamma=\frac{x+y}{c},          \nonumber \\
&&\frac{\beta}{\gamma}+\frac{\gamma}{\beta}=
\frac{1}{2bc}\left\{(\mu^{2}-\nu^{2})-(x^{2}-y^{2})\right\}   \nonumber 
\end{eqnarray}
and (\ref{eq:relations-1}) and (\ref{eq:relations-2}). Similarly, we have 
\[
2u_{11}^{2}=
\frac{2N+7+\sqrt{16(N+2)^{2}+9}}{2\sqrt{16(N+2)^{2}+9}}=
-\frac{v_{-}(N+2)}{2\sqrt{d(N+2)}}, 
\]
so that
\begin{eqnarray}
&&2u_{11}^{2}\mbox{cos}\left(tg\frac{\mu}{4}\right)+
2u_{12}^{2}\mbox{cos}\left(tg\frac{\nu}{4}\right)     \nonumber \\
=&&     
\frac{v_{+}(N+2)}{2\sqrt{d(N+2)}}
\mbox{cos}\left(tg\sqrt{\lambda_{+}(N+2)}\right)-
\frac{v_{-}(N+2)}{2\sqrt{d(N+2)}}
\mbox{cos}\left(tg\sqrt{\lambda_{-}(N+2)}\right)
=f_{2}(N+2).    \nonumber 
\end{eqnarray}

\par \noindent
The remaining 9 entries become more complicated because they contain $a$ 
and $a^{\dagger}$. See \cite{KHi} in detail.



\begin{thebibliography}{99}
%
\bibitem{TC}
M. Tavis and F. W. Cummings : 
\newblock Exact Solution for an N--Molecule--Radiation--Field Hamiltonian, 
\newblock Phys. Rev. 170(1968), 379 ; 
%
E. T. Jaynes and F. W. Cummings : 
\newblock Comparison of Quantum and Semiclassical Radiation Theories with 
Applications to the Beam Maser, 
\newblock Proc. IEEE 51(1963), 89 ; 
%
R. H. Dicke : 
\newblock Coherence in Spontaneous Radiation Processes, 
\newblock Phys. Rev, 93(1954), 99. 
%
\bibitem{books}
L. Allen and J. H. Eberly : 
\newblock Optical Resonance and Two--Level Atoms, 
\newblock Wiley, New York, 1975\ ; 
%
P. Meystre and M. Sargent III : 
\newblock Elements of Quantum Optics (third edition), 
\newblock Springer--Verlag, 1990\ ; 
%
Claude Cohen--Tannoudji, J. Dupont--Roc and G. Grynberg : 
\newblock Atom--Photon Interactions ; Basic Processes and Applications, 
\newblock Wiley, New York, 1998. 
%
\bibitem{elementary-gate}
K. Fujii : 
\newblock Introduction to Grassmann Manifolds and Quantum Computation, 
\newblock J. Applied Math, 2(2002), 371, 
\newblock quant-ph/0103011\ ; 
%
A. Barenco, C. H. Bennett, R. Cleve, D. P. Vincenzo, N. 
Margolus, P. Shor, T. Sleator, J. Smolin and H. Weinfurter : %
\newblock Elementary gates for quantum computation, 
\newblock Phys. Rev. A 52(1995), 3457, 
\newblock quant-ph/9503016. 
%
\bibitem{papers-1}
M. Orszag, R. Ramirez, J. C. Retamal and C. Saavedra : 
\newblock Quantum cooperative effects in a micromaser, 
\newblock Phys. Rev. A 49 (1994), 2933\ ; 
%
M. S. Kim, J. Lee, D. Ahn and P. L. Knight : 
\newblock Entanglement induced by a single-mode heat environment, 
\newblock Phys. Rev. A 65 (2002), 040101, 
\newblock quant-ph/0109052\ ; 
%
C. Genes, P. R. Berman and A. G. Rojo : 
\newblock Spin squeezing via atom -- cavity field coupling, 
\newblock quant-ph/0306205.
%
\bibitem{papers-2}
K. Fujii, K. Higashida, R. Kato and Y. Wada : 
\newblock Explicit Form of Solution of Two Atoms Tavis--Cummings Model, 
\newblock quant-ph/0403008\ ; 
%
K. Fujii, K. Higashida, R. Kato, T. Suzuki and Y. Wada : 
\newblock Explicit Form of the Evolution Operator of Tavis--Cummings Model : 
Three and Four Atoms Cases, 
\newblock Int. J. Geom. Methods Mod. Phys, vol.1, no.6 (2004), 721, 
\newblock quant-ph/0409068. 
%
\bibitem{FS} K. Fujii and T. Suzuki : 
\newblock A New Symmetric Expression of Weyl Ordering, 
\newblock Mod. Phys. Lett. A, 19(2004), 827, 
\newblock quant-ph/0304094.
%
\bibitem{qudit-papers}
K. Fujii : 
\newblock Exchange Gate on the Qudit Space and Fock Space, 
\newblock J. Opt. B : Quantum Semiclass. Opt, 5(2003), S613, 
\newblock quant-ph/0207002\ ; 
%
K. Fujii : 
\newblock Quantum Optical Construction of Generalized Pauli and 
Walsh--Hadamard Matrices in Three Level Systems (Lecture Note), 
\newblock quant-ph/0309132\ ; 
%
K. Fujii, K. Higashida, R. Kato and Y. Wada : 
\newblock N Level System with RWA and Analytical Solutions Revisited, 
\newblock quant-ph/0307066\ ; 
%
K. Fujii, K. Higashida, R. Kato and Y. Wada : 
\newblock A Rabi Oscillation in Four and Five Level Systems, 
\newblock quant-ph/0312060\ ; 
%
K. Funahashi : 
\newblock Explicit Construction of Controlled--U and Unitary Transformation 
in Two--Qudit, 
\newblock quant-ph/0304078. 
%
\bibitem{KHi} K. Higashida : 
\newblock Master Thesis (in preparation). 
%
\bibitem{KF-special}K. Fujii : 
\newblock A Modern Introduction to Cardano and Ferrari Formulas 
in the Algebraic Equations, 
\newblock Lecture Note, 
\newblock quant-ph/0311102. 
%
\bibitem{FHKW}K. Fujii, K. Higashida, R. Kato and Y. Wada : 
\newblock Cavity QED and Quantum Computation in the Weak Coupling Regime, 
\newblock J. Opt. B : Quantum Semiclass. Opt, 6(2004), 502, 
\newblock quant-ph/0407014. 
%
\end{thebibliography}
\end{document}